\documentclass[prd, preprint, nofootinbib, superscriptaddress
]{revtex4}

\usepackage[dvipdfmx,hiresbb]{graphicx}
\usepackage{graphicx}
\usepackage{amsfonts}
\usepackage{latexsym}
\usepackage{amsmath}
\usepackage{amssymb} 
\usepackage{epsfig}
\usepackage{bm,color}
\usepackage{slashed}
\begin{document}

\vspace*{-2cm}

\title{
SO(10) Grand Unification \\
with Minimal Dark Matter and Color Octet Scalars
}

\vspace{1cm}

\author{Gi-Chol Cho}
 \email{cho.gichol@ocha.ac.jp}
  \affiliation{
  Department of Physics, Ochanomizu University, Tokyo 112-8610, Japan
}

\author{Kana Hayami}
 \email{k.hayami@hep.phys.ocha.ac.jp}
  \affiliation{
  Graduate school of Humanities and Sciences, Ochanomizu University, Tokyo 112-8610, Japan
}

\author{Nobuchika Okada}
 \email{okadan@ua.edu}
  \affiliation{
Department of Physics and Astronomy, University of Alabama, Tuscaloosa, Alabama 35487, USA
}

%

\begin{abstract}
The minimal dark matter (MDM) scenario is a very simple framework 
   of physics beyond the Standard Model (SM) to supplement the SM with a DM candidate.  
In this paper, we consider an ultraviolet completion of the scenario to an SO(10) grand unified theory, 
   which is a well-motivated framework in light of the neutrino oscillation data. 
Considering various phenomenological constraints, 
   such as the successful SM gauge coupling unification, 
   the proton stability, and  the direct/indirect DM detection constraints 
   as well as the absolute electroweak vacuum stability, 
   we have first singled out the minimal particle content of the MDM scenario at low energies. 
 In addition to the SM particle content, our MDM scenario includes 
   an SU(2)$_L$ quintet scalar DM with a 9.4 TeV mass 
   and three degenerate color-octet scalars with mass of 2 TeV.   
We then have found a way to embed the minimal particle content into SO(10) representations,   
   in which a remnant $Z_2$ symmetry after the SO(10) symmetry breaking
    ensures the stability of the DM particle. 
The production cross section of the color-octet scalars at the Large Hadron Collider is found to be 
   a few orders of magnitude below the current experimental bound.

\end{abstract}

\preprint{OCHA-PP-368}
\vspace*{2cm}

\maketitle
\baselineskip 18pt
\section{Introduction}
\label{sec:intro}
The Large Hadron Collider (LHC) aims to understand the origin of electroweak symmetry breaking 
  and to search for physics beyond the Standard Model (SM) at the TeV scale. 
In addition to the many experimental results before the LHC, the discovery of the Higgs boson
  is another success of the SM. 
Although the SM is consistent with almost all experimental results, 
  there are some exceptions, such as the neutrino oscillation phenomena 
  and the presence of dark matter (DM) in the Universe. 
These missing pieces of the SM require us to go beyond the SM. 
One of the promising (particle) candidates of the DM is the Weekly Interacting Massive Particles (WIMP), 
  and various extensions of the SM, such as supersymmetric SMs with the $R$ parity conservation, and 
  extra dimension models with the Kaluza-Klein parity, 
  have been considered to supplement the SM with a WIMP DM candidate.  
Minimal Dark Matter (MDM) scenario \cite{Cirelli:2005uq} has been proposed as one of the simplest extensions of the SM
  to incorporate a WIMP DM to the SM. 
The scenario introduces either a fermionic or a bosonic multiplet under the SM SU(2)$_L$ gauge group,
  and its electrically neutral component in the multiplet plays a role of the DM. 
Detailed studies on the MDM scenario and its phenomenological consequences can be found in the literature
  (see Refs.~\cite{Cirelli:2007xd, Cirelli:2009uv} and papers thereof). 
As the MDM scenario is proposed as a low-energy effective theory, the stability of the MDM particle is not ensured 
  and some symmetry is introduced by hand for the DM stability.  
It is desired to consider an ultraviolet completion of the MDM scenario, 
  which can naturally provide us with a stable MDM particle at low energies. 
In this paper, we study a possibility to embed the MDM scenario into a certain grand unified theory (GUT), 
  in particular, an SO(10) GUT. 
The models based on the SO(10)  gauge group is well-motivated in the light of the neutrino oscillation data: 
One right-handed neutrino is unified with the SM chiral fermions in each generation 
  to form a {\bf 16}-plet fermion under the SO(10), 
  and the seesaw mechanism \cite{Minkowski:1977sc, Yanagida:1979as, Gell-Mann:1979vob, Mohapatra:1979ia}  
  naturally explaining the tiny neutrino masses is automatically implemented 
  in association with the SO(10) symmetry breaking down to the SM gauge groups. 
As a low energy effective theory, we introduce an MDM multiplet (either fermionic or bosonic) to the SM particle content. 
In addition to the MDM multiplet, we find that the presence of SU(3)$_C$ multiplets (but neutral under SU(2)$_L \times$U(1)$_Y$)
  at the TeV scale is crucial to realize a successful gauge coupling unification. 
Taking into account the gauge coupling unification (below the Planck scale) and the constraint on the proton lifetime, 
  we determine the representation of MDM and colored new particles. 
We then discuss how the selected MDM and colored multiplets can be embedded into some SO(10) multiplets
  to ensure the stability of the MDM particle by a remnant symmetry of SO(10) breaking \cite{Frigerio}.
We also show that in the presence of the selected MDM and colored particles, 
  the SM Higgs quartic coupling remains positive in its renormalization group (RG) evolution below the Planck (or GUT) scale, 
  so that the instability issue of the electroweak vacuum is resolved. 
The existence of the new colored particles has no conflict with the current LHC data. 
%

%
This paper is organized as follows:
In Sec.~\ref{sec:mdmreview}, we briefly review the MDM scenario, and then select the representation
   of the MDM multiplet so as to satisfy the experimental bound on the proton lifetime. 
In order to achieve a successful unification of the three SM gauge couplings below the Planck scale, 
   we introduce SU(3)$_C$ multiplet scalars in Sec.~\ref{sec:color_GUT}. 
The best choice of SU(3)$_C$ representation is discussed in terms of the absolute stability for the electroweak vacuum. 
We also discuss the current LHC constraints on the colored scalars from the search for a narrow resonance 
   with dijet final states. 
In Sec.~\ref{sec:GUT_embdedding}, we consider the embedding of our MDM scenario into an SO(10) GUT and find the smallest 
   representation which ensures the MDM particle stability by a remnant parity after the SO(10) symmetry breaking. 
Sec.~\ref{sec:conclusion} is devoted to conclusions.

\section{MDM scenario and Selection of Representation}
\label{sec:mdmreview}

In the MDM scenario, an SU(2)$_L$ multiplet $\chi$ is introduced to the SM particle content. 
With an appropriate assignment of a hypercharge $Y_\chi$ for $\chi$, 
  the multiplet includes an electrically neutral component which plays a role of DM.\footnote{
Unless some confusion arises, we use the same symbol $\chi$ for both the MDM multiplet and the DM particle
throughout this paper.}
The kinetic and mass parts in the Lagrangian of the multiplet $\chi$ are given by \cite{Cirelli:2005uq,  Cirelli:2007xd, Cirelli:2009uv}
\begin{align}
\mathcal{L}_{\mathrm{kin}}
=
c
\begin{cases}
  \overline{\chi} (i \gamma^\mu D_\mu -M) \chi
  & \mbox{for a fermionic $\chi$},
  \\
  \left|D_{\mu} \chi \right|^{2} - M^{2}|\chi|^{2}
  & \mbox{ for a scalar $\chi$},
\end{cases}
\label{eq:lagkin}
\end{align}
where $c=1/2$ or $1$ depending on the representation of the SU(2)$_L$ multiplet $\chi$ being real or complex, 
  $D_{\mu}$ represents the covariant derivative with respect to $\mathrm{SU(2)}_L \times \mathrm{U(1)}_Y$,
  and $M$ is the mass of $\chi$.
Although we can consider $Y_\chi \neq 0$ in general, we set $Y_\chi=0$ in this paper. 
This is because if $Y_\chi \neq 0$ an MDM particle can scatter off a nucleon
  through the $Z$-boson exchange process while this spin-independent DM scattering cross section
  is very severely constrained by the direct DM detection experiments. 
For example, the XENON1T~\cite{Aprile:2018dbl} experiment excludes $M_\chi \lesssim10^9$ GeV
  for the mass of a WIMP DM with non-zero hypercharge (see, for example, Refs.~\cite{Dunsky:2020yhv, Okada:2021uqk}). 
For a WIMP DM with a large mass avoiding the XENON1T constraint, 
  the standard freeze-out mechanism leads to the relic DM density being overabundant \cite{Griest:1989wd}.  
Note that the stability of an MDM particle is not ensured by the SM gauge invariance, 
  and therefore some symmetry should be introduced to prevent the MDM to decay. 
Once the representation of $\chi$ under the SU(2)$_L$ is fixed, the DM mass $M$ is the only free parameter
  in the MDM scenario. 
This DM mass is determined so as to reproduce the observed DM relic density 
  from the precise measurements of the cosmic microwave background~\cite{Aghanim:2018eyx}:
\begin{align}
\Omega_\chi h^2 = 0.120 \pm 0.001. 
\label{eq:DMrelic}
\end{align}
In Table~\ref{tab:1} (the second column), we summarize the $M$ values for various representations of the MDM 
\cite{Cirelli:2005uq,  Cirelli:2007xd, Cirelli:2009uv}.\footnote{
See Ref.~\cite{Bottaro:2021snn} for a very recent update of the MDM masses 
  from precise computations including Sommerfeld enhancement and bound states formation
  at leading order in the SM gauge boson exchange and emission.
The corrections of the MDM masses have negligible effects on our renormalization group analysis.    
}
\footnote{
Note that a scalar MDM can generally have a gauge invariant coupling with the SM Higgs doublet $H$, 
  such as $\lambda_{H \chi} \chi^\dagger  \chi H^\dagger H$. 
Through this Higgs-portal coupling, a pair of MDMs can annihilate to the Higgs doublets with a cross section 
  approximately given by $\sigma (\chi^\dagger \chi \to H^\dagger H) \sim \frac{\lambda_{H \chi}^2}{4 \pi M^2} $
  for $M \gtrsim 1$ TeV in the non-relativistic limit. 
We assume that $\lambda_{H \chi} \ll g_2^2$, where $g_2$ is the SU(2)$_L$ gauge coupling, 
  so that the high predictability of the MDM scenario is kept intact.  
}

\begin{table}[t]
\begin{center}
\begin{tabular}{ |c | c | c | c | c | c | c | c | }
\hline
\hline
\multicolumn{7}{c}{Scalar MDM} \\
\hline
\hline
Rep. in SU(2)$_L$ & $M$ [TeV] & $\Delta b_2$ & $ \Delta b_{22}$ &  $M_X$ [GeV] & $\alpha_X$ & $\tau_p$ [yr] & Viable? \\
\hline
{\bf 3} (real) & 2.5 & $\frac{1}{3}$ & $ \frac{16}{3}$ &  $ 3.6 \times 10^{13}$ & $1/42$ & $ 4.5 \times 10^{25} $ & No \\
{\bf 3} (complex)  & 2.5 & $\frac{2}{3}$ & $ \frac{56}{3}$ &  $ 1.3 \times 10^{14}$ & $1/41$ & $ 7.7 \times 10^{27} $ & No \\
{\bf 5} (real) & 9.4 & $\frac{5}{3}$ & $ \frac{200}{3}$ &  $ 1.3 \times 10^{16}$ & $1/38$ &$ 6.4 \times 10^{35} $ & Yes \\
{\bf 5} (complex) & 9.4 & $\frac{10}{3}$ & $ \frac{760}{3}$ &  $ > M_P$ & $-$ &$ - $ & No \\
{\bf 7} (real) & 25 & $\frac{14}{3}$ & $ \frac{1064}{3}$ &  $ > M_P$ & $-$ & $ - $ & No \\
\hline
\hline
\multicolumn{7}{c}{Fermion MDM} \\
\hline
\hline
Rep. in SU(2)$_L$ & $M$ [TeV] & $\Delta b_2$ & $ \Delta b_{22}$ &  $M_X$ [GeV] & $\alpha_X$ & $\tau_p$ [yr] & Viable? \\
\hline
{\bf 3} & 2.7 & $\frac{4}{3}$ & $ \frac{64}{3}$ &  $ 2.2 \times 10^{15}$ & $1/39$ & $ 5.4 \times 10^{32} $ & Marginal \\
{\bf 5} & 10 & $\frac{20}{3}$ & $ \frac{560}{3}$ &  No Merger  & $-$ & $ - $ & No \\
\hline
\hline
\end{tabular}
\end{center}
\caption{
Candidates of the MDM multiplets are shown.
Upper-five and lower-two cases represent the scalar and fermion DMs, respectively.
DM masses which reproduce the observed relic density are shown in the second column.
The coefficients of beta function for SU(2)$_L$ gauge coupling in the 1- and 2-loop levels are given in the third and fourth columns.
The fifth column shows the energy where two gauge couplings, $\alpha_1$ and $\alpha_2$, merge, 
   and the unified gauge coupling values are in the sixth column. 
The predicted proton lifetime is listed in the seventh column. 
The last column shows the consistency with the experimental lower bound on the proton lifetime. 
}
\label{tab:1}
\end{table}
%

%
Let us now examine the unification of the SM gauge couplings with the MDM multiplets at the TeV scale. 
Since the MDM multiplet $\chi$ has the SU(2)$_L$ charge, 
   the RG evolution of the SU(2)$_L$ gauge coupling is altered. 
We define the gauge coupling unification scale  $M_X$ at which the SU(2)$_L$ and U(1)$_Y$ gauge couplings merge. 
Clearly, the SU(3)$_C$ gauge coupling is not unified with the others at $M_X$. 
As we will discuss in the next section, the introduction of colored particles (but neutral under SU(2)$_L \times $ U(1)$_Y$) 
   plays a crucial role in the three SM gauge coupling unification. 
In our study on the gauge coupling unification, we employ the RG equations at the 2-loop level \cite{RGE1, RGE2, RGE3, RGE4, RGE5}. 
In the SM, we have 
\begin{eqnarray}
 \mu \frac{d g_i}{d \mu} =
 \frac{b_i}{16 \pi^2} g_i^3 +\frac{g_i^3}{(16\pi^2)^2}
  \left( \sum_{j=1}^3 b_{ij}g_j^2 - c_i y_t^2   \right),
\label{eq:rge}  
\end{eqnarray}
 where $\mu$ is the renormalization scale, $g_i$ ($i=1,2,3$) correspond to the SM three gauge couplings 
  with the SU(5) GUT normalization for $g_1 = \sqrt{5/3} \, g_Y$, and  
\begin{align}
  b_i = \left(
  \begin{array}{c}
  41/10 \\
  -19/6 \\
  -7
  \end{array}
  \right),
~~
  b_{ij} =
  	\left(
    \begin{array}{ccc}
  	199/50 & 27/10 & 44/5 \\
  	9/10 & 35/6 & 12 \\
  	11/10& 9/2  & -26
  	\end{array}
    \right), 
~~   
  c_i = \left(
  \begin{array}{c}
  17/10 \\
   3/2 \\
   2
  \end{array}
  \right). 
    \label{eq:coeffsm}
\end{align}
Here, among the SM Yukawa couplings, we include only the top Yukawa coupling ($y_t$). 
The RG equation for the top Yukawa coupling given by
\begin{eqnarray} 
\mu  \frac{d y_t}{d \mu}
 = y_t  \left(
 \frac{1}{16 \pi^2} \beta_t^{(1)} + \frac{1}{(16 \pi^2)^2} \beta_t^{(2)}
 \right), 
\end{eqnarray}
where the one-loop contribution is
\begin{eqnarray}
 \beta_t^{(1)} =  \frac{9}{2} y_t^2 -
  \left(
    \frac{17}{20} g_1^2 + \frac{9}{4} g_2^2 + 8 g_3^2
  \right) ,
\end{eqnarray}
while the two-loop contribution is 
\begin{eqnarray}
\beta_t^{(2)} &=&
 -12 y_t^4 +   \left(
    \frac{393}{80} g_1^2 + \frac{225}{16} g_2^2  + 36 g_3^2
   \right)  y_t^2  \nonumber \\
 &&+ \frac{1187}{600} g_1^4 - \frac{9}{20} g_1^2 g_2^2 +
  \frac{19}{15} g_1^2 g_3^2
  - \frac{23}{4}  g_2^4  + 9  g_2^2 g_3^2  - 108 g_3^4 \nonumber \\
 &&+ \frac{3}{2} \lambda^2 - 6 \lambda y_t^2 .
\end{eqnarray}
The RG equation for the quartic Higgs coupling is given by 
\begin{eqnarray}
\mu \frac{d \lambda}{d \mu}
 =   \frac{1}{16 \pi^2} \beta_\lambda^{(1)}
   + \frac{1}{(16 \pi^2)^2}  \beta_\lambda^{(2)},
\end{eqnarray}
with
\begin{eqnarray}
 \beta_\lambda^{(1)} &=& 12 \lambda^2 -
 \left(  \frac{9}{5} g_1^2+ 9 g_2^2  \right) \lambda
 + \frac{9}{4}  \left(
 \frac{3}{25} g_1^4 + \frac{2}{5} g_1^2 g_2^2 +g_2^4
 \right) + 12 y_t^2 \lambda  - 12 y_t^4 ,
\label{lam_1}
\end{eqnarray}
at the 1-loop level and
\begin{eqnarray}
  \beta_\lambda^{(2)} &=&
 -78 \lambda^3  + 18 \left( \frac{3}{5} g_1^2 + 3 g_2^2 \right) \lambda^2
 - \left( \frac{73}{8} g_2^4  - \frac{117}{20} g_1^2 g_2^2
 - \frac{1887}{200} g_1^4  \right) \lambda - 3 \lambda y_t^4
 \nonumber \\
 &&+ \frac{305}{8} g_2^6 - \frac{289}{40} g_1^2 g_2^4
 - \frac{1677}{200} g_1^4 g_2^2 - \frac{3411}{1000} g_1^6
 - 64 g_3^2 y_t^4 - \frac{16}{5} g_1^2 y_t^4
 - \frac{9}{2} g_2^4 y_t^2
 \nonumber \\
 && + 10 \lambda \left(
  \frac{17}{20} g_1^2 + \frac{9}{4} g_2^2 + 8 g_3^2 \right) y_t^2
 -\frac{3}{5} g_1^2 \left(\frac{57}{10} g_1^2 - 21 g_2^2 \right)
  y_t^2  - 72 \lambda^2 y_t^2  + 60 y_t^6,
\end{eqnarray}
at the 2-loop level. 
In solving the RG equations, we use the boundary conditions at the top quark pole mass ($m_t$)
  given in Ref.~\cite{RGE_Higgs_quartic}: 
$g_1(m_t)= 0.46256$, $g_2(m_t)= 0.64779$, $g_3(m_t)= 1.1666$, $y_t(m_t) = 0.9369$, and $\lambda(m_t) = 0.25183$, 
  for the choice of  $\alpha_s =0.1184$, $m_W=80.384$ GeV, and $m_t=173.34$ GeV. 
For $\mu >M$, the contributions from MDM multiplet to the beta function coefficients, 
  $\Delta b_2$ and $\Delta b_{22}$ at the 1- and 2-loop levels, respectively, should be added. 
These values for various MDM multiplets are listed in Table~\ref{tab:1} (the third and forth columns). 
The GUT scales ($M_X$) for the various MDM multiplets and the unified gauge coupling ($\alpha_X$)
  are listed in Table~\ref{tab:1} (the fifth and sixth columns).

The proton decay mediated by the GUT gauge bosons is a characteristic prediction of the GUTs. 
The current lower limit of the proton lifetime has been set by the super-Kamiokande experiment~\cite{Nishino:2009aa}: 
\begin{align}
\tau (p \rightarrow \pi^{0} e^{+} )_\mathrm{exp}
>
 8.2 \times 10^{33} \, \mathrm{yrs}. 
\end{align}
Once $M_X$ and $\alpha_X=\frac{g_X^2}{4 \pi}$ are fixed by the RG analysis, 
  we can estimate the proton lifetime. 
For the proton decay processes mediated by the GUT gauge bosons with mass $M_X$, 
  we estimate the proton lifetime as \cite{Frigerio}\footnote{
For simplicity, we here consider the proton decay process mediated by the SU(5) GUT gauge boson $X$ 
  of the representation $({\bf 3}, {\bf 2}, -5/6)$ under the SM gauge groups of SU(3)$_C \times$SU(2)$_L \times$U(1)$_Y$. 
The lifetime is reduced by a factor $4/5$ if we add the processes mediated by the other SU(5) gauge boson  
  $Y: ({\bf 3}, {\bf 2}, 1/6)$  \cite{Frigerio}. 
}
\begin{align}
	\tau(p \rightarrow \pi^{0} e^{+})_\mathrm{th}
  \simeq
  \left(
  8.2 \times 10^{33} \, \mathrm{yrs}
  \right)
  \left(
  \frac{1 / 39}{\alpha_{\mathrm{X}}
  }\right)^2
  \left(
  \frac{M_{X}}{4.3 \times 10^{15}~\mathrm{GeV}}
  \right)^4.
  \label{protonDecay}
\end{align}
With the resultant $M_X$ and $\alpha_X$ from the RG analysis, 
   we list the predicted proton lifetime for various MDM multiplets in Table~\ref{tab:1} (the seventh column). 
For the scalar MDM scenario, the real scalar quintet satisfies all the constraints. 
On the other hand, no definitely viable case is found for the fermionic MDM case. 
However, if the threshold corrections from a large SO(10) multiplet,  
  which the MDM multiplet is embedded into, are taken into account, 
  we may consider a theoretical uncertainty of the proton lifetime to be an order of magnitude.  
Hence we have marked the triplet fermion MDM as ``Marginal" in Table~\ref{tab:1}.
See Ref.~\cite{Frigerio} for a detailed study on an SO(10) GUT realization of this triplet fermion MDM scenario. 

With the requirement of the gauge coupling unification below the Planck scale and the constraint on the proton lifetime, 
  we have selected two phenomenologically viable cases for the MDM scenario: 
  the real-scalar quintet and the fermion triplet.   
Let us here consider the constraints on these two MDMs from the direct and indirect DM search experiments. 
Although we set $Y_\chi=0$ to avoid the MDM particle to scatter off with nucleon
  through $Z$-boson exchange at the tree-level, scattering processes at the quantum level
  may give rise to a sizable contribution to the spin-independent cross section $\sigma^p_{\mathrm{SI}}$.
In Ref.~\cite{Hisano:2015rsa}, the authors evaluated the next-to-leading order contributions
  to $\sigma^p_{\mathrm{SI}}$ for a generic SU(2)$_L$ multiplet DM
  with a vanishing hypercharge. 
The cross sections for SU(2)$_L$ triplet and quintet fermion DMs with a mass of 1 TeV 
   are found to be $\sigma^p_{\mathrm{SI}} \simeq 10^{-47}$ cm$^2$ 
   and $10^{-49}$ cm$^2$, respectively,
   which are a few orders of magnitude below the current experimental upper bound 
   set by the XENON1T experiment \cite{Aprile:2018dbl}.
Although the scalar DM case has not been calculated in Ref.~\cite{Hisano:2015rsa}, 
   we expect that the resultant cross section for the quintet scalar DM will be in the same order
   as the one for the quintet fermion DM. 
Therefore, we conclude that our two cases are consistent with the current direct DM search experiments. 
Indirect DM detection experiments have been searching for an excess of cosmic-rays
  originating from the annihilations and/or decays of DM particles in the halo of our galaxy. 
In Ref.~\cite{Cuoco:2017iax}, the MDM annihilation cross sections for all possible channels into 
  pairs of SM particles are calculated, and it has been found that the cosmic-ray antiproton data 
  observed by AMS-02~\cite{Aguilar:2016kjl} provide the constraints on the MDM annihilation cross sections
  more stringent than that derived from gamma-ray observations. 
In particular, it has been concluded in Ref.~\cite{Cuoco:2017iax} that the SU(2)$_L$ triplet fermion MDM is strongly disfavored. 
The quintet fermion MDM is also severely constrained but is still consistent with the antiproton data. 
Concerning these results, we focus on the quintet scalar MDM in the following.\footnote{
Although the annihilation cross section of the quintet scalar MDM is not calculated in Ref.~\cite{Cuoco:2017iax},
  we expect the cross section to be the same order of magnitude as the one of the fermion quintet DM. 
}

\section{Selection of colored multiplet}
\label{sec:color_GUT}
As mentioned in the previous section, we consider the presence of SU(3)$_C$ multiplets
  but neutral under SU(2)$_L \times $U(1)$_Y$ for the successful unification of 
  the three SM gauge couplings. 
While such SU(3)$_C$ multiplets should be unstable, they have no couplings with the SM particles 
  at the renormalizable level. 
As a simple possibility, we consider scalar fields (we call it $\phi$ in the following) 
  of $\bm{8},\bm{10}$ and $\bm{27}$ representations under SU(3)$_C$, 
  which have couplings with two gluons through a gauge invariant dimension-5 operator. 
This operator is induced by quantum corrections at the 1-loop level (see Appendix). 
For a fixed representation ($\bm{8},\bm{10}$ or $\bm{27}$),  
   we introduce $N$ scalars with a degenerate mass $m_\phi$
   and analyze the RG evolution of the SU(3)$_C$ coupling $g_3$ at the two-loop level.  
The scalar mass $m_\phi$ is determined so as to realize the SM gauge coupling unification
   at $M_X=1.3 \times 10^{16}$ GeV, 
   which has been determined by $g_1(M_X)=g_2(M_X)$
   in the presence of the real scalar quintet MDM (see Table~\ref{tab:1}).\footnote{
In precise, $M_X$ depends on the colored particle choice,
   since the RGEs of $g_{1,2,3}$ at the 2-loop level involve $y_t$. 
We find that the change of $M_X$ is very small (less than 1\%). 
}       
In Table~\ref{tab:2}, we show our results for various choices for the representation and $N$. 
For $\mu \geq m_\phi$, the contributions of colored scalars to the beta functions at 1- and 2-loop level 
   are denoted as $\Delta b_3$ and $\Delta b_{33}$, respectively. 
We have found that with $N=1, 2$ octet scalars, 
   the gauge coupling unification cannot be achieved for $m_\phi \geq m_t$. 
For a given representation, the resultant $m_\phi$ is larger for a larger $N$ value. 
%
\begin{table}[t]
\begin{center}
\begin{tabular}{|c|c|c|c|c|c|}
\hline
\hline
Rep. in SU(3)$_C$ & $N$ & $m_\phi$ [TeV] & $\Delta b_3$ & $ \Delta b_{33}$ &  Vacuum stability \\
\hline
{\bf 8}     & $3$ & $2.0 $ & $\frac{3}{2}$ & $ 36 $ &   Yes   \\
                    & 4 & $2.5 \times 10^3$ & $ 2 $ & $ 48 $ &   No  \\
                    & 5 & $ 2. 0 \times 10^5 $ & $\frac{5}{2}$ & $ 60 $  & No  \\
\hline
{\bf 10}        & 1 & $1.5 \times 10^6 $  & $\frac{5}{2}$ & $ 195 $ &No  \\
                   & 2 & $3.7 \times 10^9 $  & $\frac{10}{2}$ & $ 390 $ & No  \\
\hline
{\bf 27}        & 1 & $1.6 \times 10^{11} $  & $ 9 $ & $ 918 $ & No  \\
                   & 2 & $1.4 \times 10^{12} $  & $ 18 $ & $ 1836 $ &  No  \\
\hline
\hline

\end{tabular}
\end{center}
\caption{
Candidates of scalar SU(3)$_C$ multiplets.
}
\label{tab:2}
\end{table}

%
\begin{figure}[htbp]
\centering
\includegraphics[clip,width=7.7cm]{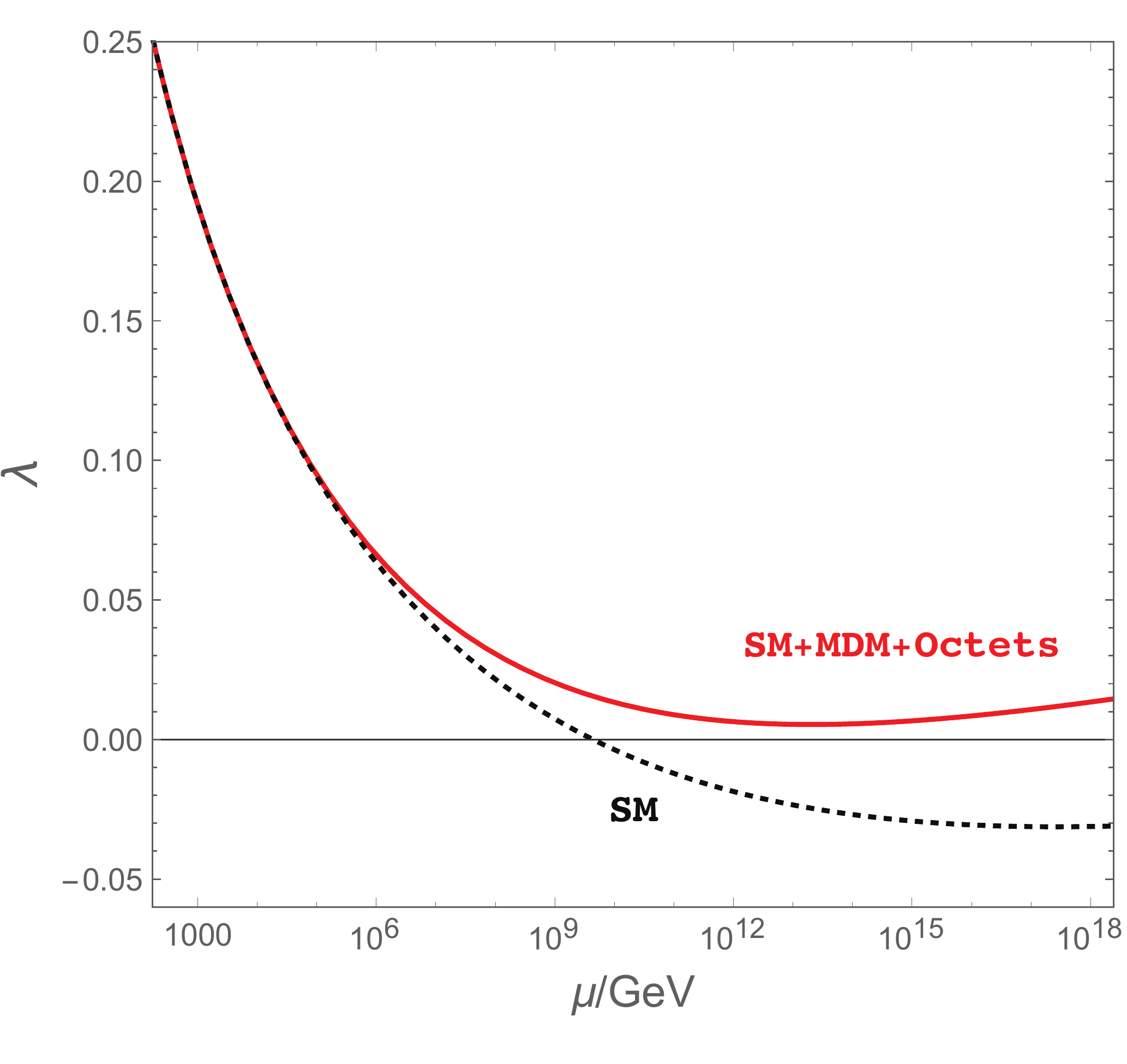}\; \; \;
\includegraphics[clip,width=7.5cm]{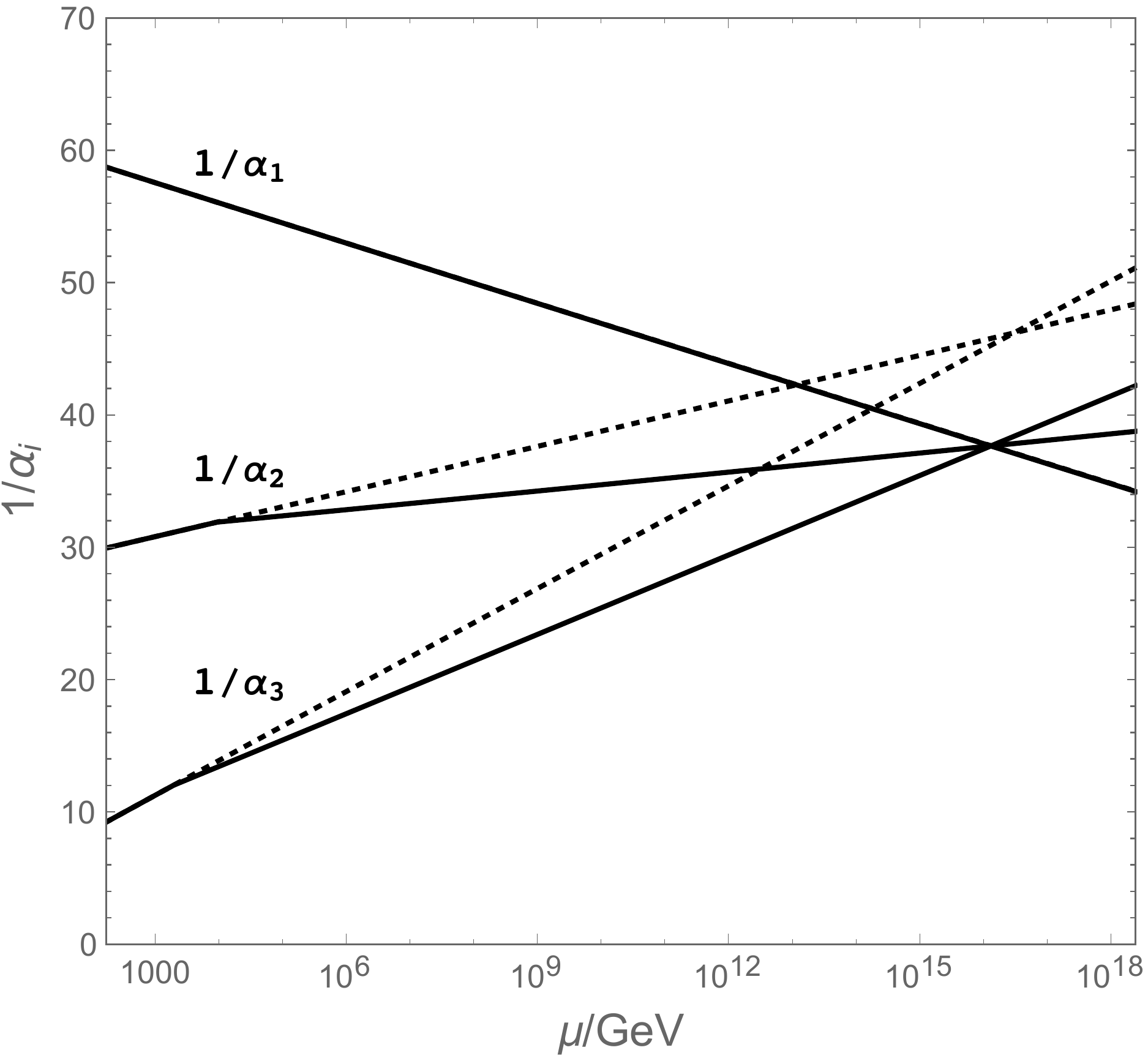}
\caption{
{\it Left Panel}: The RG evolutions of the SM Higgs quartic coupling with the SM particle content (dashed line)
 and the SM particle content plus SU(2)$_L$ ${\bf 5}$-plet real scalar DM and
3 SU(3)$_c$ octet real scalars (red line).
{\it Right Panel}: The evolution of the gauge coupling evolutions (solid lines), 
   along with the SM case results (dashed lines).
}
\label{Fig:sigmav:indirect}
\end{figure}

%
In order to find the best choice of $\phi$ from the list in Table~\ref{tab:2}, 
  we may consider the absolute stability of the electroweak vacuum. 
It is known that with the SM particle content, the Higgs quartic coupling ($\lambda$) becomes negative 
  in its RG evolution at $\mu \simeq 10^{10}$ GeV \cite{RGE_Higgs_quartic}, 
  which indicates that the electroweak vacuum is unstable. 
In the presence of the quintet MDM and the colored particles, 
  the RG evolution of $\lambda$ is altered, so that there is a possibility 
  for $\lambda(\mu)$ to remain positive up to the Planck (or GUT) scale. 
In general, we can introduce mixed quartic couplings of the Higgs doublet 
  with the scalar MDM and the colored scalars, such as 
\begin{eqnarray}
\mathcal{L}
\supset
\lambda_{H\phi}
\left( H^\dagger H \right) \mathrm{Tr}[\phi^2],
\end{eqnarray}
where $H$ is the SM Higgs doublet. 
If this mixed quartic coupling $\lambda_{H\phi}$ is sizable, 
  its contribution to the beta function of $\lambda$, 
  which is proportional to $\lambda_{H\phi}^2$ at the 1-loop level, 
  can make $\beta_\lambda^{(1)} >0$  and hence, 
  the absolute stability of the electroweak vacuum is maintained. 
Here, let us assume that such mixed quartic couplings are negligibly small for simplicity.\footnote{
Similar to footnote 3, 
 the quintet scalar MDM can have a gauge invariant coupling with the colored scalar, 
 $\lambda_{\phi \chi} \chi \chi \mathrm{Tr}[\phi^2]$. 
Through this coupling, a pair of quintet MDMs can annihilate to a pair of the colored scalars 
  with a cross section approximately given by $\sigma (\chi \chi \to \phi \phi) \sim \frac{\lambda_{\phi \chi}^2}{4 \pi M^2} $
  for $M > m_\phi$ in the non-relativistic limit.
We assume that $\lambda_{\phi \chi} \ll g_2^2$, where $g_2$ is the SU(2)$_L$ gauge coupling, 
  so that the high predictability of the MDM scenario is kept intact 
  and the contribution of $\lambda_{\phi \chi}$ to the beta function $\lambda$ is negligible. 
}
Note that even without the new couplings, $\beta_\lambda$ can turn to be positive 
  at high energy in the presence of the quintet MDM and the colored particles. 
This can be understood as follows: 
In the presence of the quintet MDM, $g_2(\mu)$ is larger than its value in the SM 
  for $\mu \geq M$ and thus $\beta_\lambda^{(1)}$ receives more positive contributions
  from $g_2^4$ (see Eq.~(\ref{lam_1})). 
In the presence of the colored multiplets, 
  $g_3(\mu)$ is larger than its value in the SM for $\mu \geq m_\phi$.
The 1-loop beta function of $y_t$ indicates that the running $y_t$ reduces faster than the SM case 
  toward high energy, which means the negative contribution of $y_t^4$ to $\beta_\lambda^{(1)}$ 
  becomes milder.

With the MDM and the colored multiplets listed in Table~\ref{tab:2}, 
  we calculate the evolution of $\lambda$ at the two-loop level 
  to see if $\lambda$ remains positive below the Planck (or GUT) scale. 
We find that only the case with the $N=3$ color-octet scalars can satisfy the absolute vacuum stability condition. 
Now our MDM scenario that can satisfy all the requirements is defined 
  by the particle content: the SM particles + one scalar quintet MDM with a mass 9.4 TeV + 3 degenerate color-octet scalars with the mass of 2 TeV. 
The RG evolution of $\lambda$ for this case is shown in Fig.~\ref{Fig:sigmav:indirect} (left panel),
  along with the RG evolution in the SM case.  
We also show in the right panel the evolutions of the three SM gauge couplings in the MDM scenario (solid lines), 
   along with those in the SM (dashed lines). 
The three SM gauge couplings are successfully unified at $M_X=1.3 \times 10^{16}$ GeV.

Our MDM scenario includes the 3 color-octet scalars with the mass of 2 TeV. 
They can be produced at the LHC through gluon fusion and subsequently decay into two gluons. 
Here we briefly discuss the possibility to detect the color-octet scalars at the LHC. 
Neglecting $\lambda_{H\phi}$, the Lagrangian involving one octet scalar $\phi$ is given by 
\begin{align}
\mathcal{L}
=
  \mathrm{Tr} | D_{\mu} \hat{\phi} |^2
- m^2_{\phi}\mathrm{Tr}  \left[\hat{\phi}^2 \right]
- \lambda_3 \mathrm{Tr}  \left[\hat{\phi}^3 \right]
- \lambda_4 \mathrm{Tr} \left[\hat{\phi}^4 \right] ,
\label{L_octet}
\end{align}
where $\hat{\phi} \equiv \phi^a T^a$ ($a=1,2,3$) with $T^a$ being a generator of $\mathrm{SU(3)}_C$, 
   $\lambda_3$ is a cubic coupling with mass-dimension one, 
   and $\lambda_4 > 0$ is a dimensionless quartic coupling. 
For simplicity, we only consider three copies of the Lagrangian for three octet scalars, 
   although, in general, we can introduce mixed cubic and quartic couplings among the three octets. 
Note that the non-zero cubic coupling is necessary to induce the effective coupling of $\phi$ with a pair of gluons 
   through quantum corrections at the 1-loop level. 
Following Ref.~\cite{Han:2010rf}, we parametrize the operator as
\begin{eqnarray}
    {\cal L}_{eff} = g_3 \, d^{a b c} \, \frac{\kappa_\phi}{\Lambda_\phi}  \phi^a \, G_{\mu \nu}^b \,G^{c, \mu \nu} ,%
\label{dim-5}
\end{eqnarray}   
   where $G_{\mu \nu}^a$ is the gluon field-strength, $\kappa_\phi$ is a constant,
   $\Lambda_\phi$ is a mass parameter, and $d^{a b c}$ is a constant determined by the SU(3) algebra, 
   $\left\{ T^a, T^b  \right\} = \frac{1}{3}\delta^{a b} + d^{a b c} T^c$.
As shown in Appendix, the parameters $\kappa_\phi$ and $\Lambda_\phi$ are determined by 1-loop calculations. 
Note that we can introduce an upper bound on the cubic coupling $|\lambda_3|$ not to generate 
   a colored breaking minimum in the octet scalar potential. 
For example, if we consider a direction of the vacuum expectation value $\langle \phi \rangle = \langle \phi^8 \rangle \, T^8$ 
   with $T^8=diag(1, 1,-2)/\sqrt{12}$, we find an upper bound $|\lambda_3| < 4 \sqrt{\frac{2 \lambda_4}{3}} m_\phi$
   to avoid a color breaking (local) minimum in the octet scalar potential. 
For a perturbative value of $\lambda_4 \lesssim 1$, we have $|\lambda_3|/m_\phi \lesssim 1$.

A high-mass narrow resonance decaying to a pair of jets has been searched at the LHC Run-2. 
The CMS collaboration has reported the result of their analysis with an integrated luminosity of 137/fb 
  \cite{CMS:2019gwf}.  
The upper bound on the product of the cross section ($\sigma(pp \to \phi)$),
  the branching fraction ($B$), and the acceptance ($A$) has been set as
  $\sigma B A \lesssim 0.2$ pb for $m_\phi=2$ TeV, which should be compared to the theoretical prediction, 
  $\sigma B A \simeq 4$ pb, from the dimension-5 operator of Eq.~(\ref{dim-5}) 
  with $\kappa_\phi=1/\sqrt{2}$ and $\Lambda_\phi=m_\phi$ \cite{CMS:2019gwf}. 
In Appendix, the parameters involved in the dimension-5 operator are given in terms of $g_3$, $\lambda_3$, and $m_\phi$ through 1-loop calculations. 
The relation is roughly given by 
\begin{eqnarray}
   g_3 \frac{\kappa_\phi}{\Lambda_\phi} \sim \frac{\lambda_3 g_3^2}{16 \pi^2 m_\phi^2} \; 
   \to \;
  \kappa_\phi \sim 
  \frac{g_3}{16 \pi^2}  \left( \frac{\lambda_3}{m_\phi} \right) \left( \frac{\Lambda_\phi}{m_\phi} \right),  
  \end{eqnarray}    
where the factor $\frac{1}{16 \pi^2}$ is from the 1-loop corrections. 
Since the production cross section of $\phi$ is proportional to $\kappa_\phi^2$, 
  the theory prediction $\sigma B A \simeq 4$ pb with $\kappa_\phi=1/\sqrt{2}$ and $\Lambda_\phi=m_\phi$
  is scaled by a factor $\frac{\alpha_3}{32 \pi^3} \left( \frac{\lambda_3}{m_\phi} \right)^2 
  \simeq 1.2 \times10^{-4}$ for $\lambda_3/m_\phi = 1$.  
Taking the contribution from three identical octet scalars into account,
  we conclude that $\sigma B A$ in our model is about 2 orders of magnitude smaller than the current CMS limit.

\section{SO(10) GUT embedding}
\label{sec:GUT_embdedding}
\begin{table}[t]
	\begin{center}
		\begin{tabular}{c|c|c|c|c} \hline \hline
			SO(10) & $\bm{660}$ & $\bm{770}$ & $\bm{1386}$ & $\bm{2640}$
			\\ \hline
			$\mathrm{SU(5)}\times \mathrm{U(1)}_X$
			& $(\bm{200},0)$ & $(\bm{200},0)$ & $(\bm{200},0)$ & $(\bm{200},5)$
		 \\ \hline \hline
		\end{tabular}
		\caption{
Representations in SO(10) which contains $\bm{200}$ in SU(5).
The $\mathrm{U(1)}_X$ charge of each $\bm{200}$ is also shown.
		}
		\label{representation}
	\end{center}
\end{table}

%
In the MDM scenario, the stability of the MDM is not automatically ensured,
   and we assume some symmetry to prevent the DM particle to decay. 
In this section, we consider an embedding of our MDM scenario into an SO(10) GUT,
   in which the MDM stability is ensured due to a remnant symmetry after SO(10) breaking.

Since the rank of SO(10) is five, there is an extra U(1) gauge symmetry (which we call U(1)$_X$) 
   orthogonal to the SM gauge symmetry $G_{\mathrm{SM}}=\mathrm{SU(3)}_C \times \mathrm{SU(2)}_L \times \mathrm{U(1)}_Y$, 
   and the SO(10) group includes $G_{\mathrm{SM}} \times \mathrm{U(1)}_X$ as its subgroup. 
If the $\mathrm{U(1)}_X$ symmetry is broken by a vacuum expectation value of a Higgs field 
   with an even U(1)$_X$ charge, $Z_2$ symmetry arises as a remnant symmetry.
Then, the parity of a field $F$ with a U(1)$_X$ charge $Q_F$ is defined as $P=(-1)^{Q_F}$. 
Note that the SM fermions (plus one right-handed neutrino) in each generation 
   belong to ${\bf 16}$-plet under SO(10) with an odd parity $P=-1$
   while the SM Higgs doublet is embedded into ${\bf 10}$-plet with $P=+1$. 
The parity for the SM gauge bosons is also even. 
Therefore, if the MDM particle in our model, which is an SU(2)$_L$ quintet scalar, 
   has an odd parity, its decay to the SM particles is forbidden by the parity conservation
   and the Lorentz invariance. %

Let us now consider the embedding of $G_{\mathrm{SM}}$ into the SU(5) subgroup of SO(10) 
   such that SO(10)$\supset$SU(5)$\times$U(1)$_X \supset G_{\mathrm{SM}} \times \mathrm{U(1)}_X$.  
We find that the smallest representation of SU(5) which contains the scalar quintet MDM $(\bm{1}, \bm{5}, 0)$ 
   under $G_{\mathrm{SM}}$ is $\bm{200}$ \cite{Yamatsu:2015npn}.  
There are several representations in SO(10) which have the $\bm{200}$ representation in its SU(5) subgroup 
   as shown in Table~\ref{representation}~\cite{Yamatsu:2015npn}. 
We find that the smallest representation with an odd U(1)$_X$ charge is $\bm{2640}$.
The representation of the color-octet scalar $\phi$ is the same as SU(3)$_C$ gauge boson with $P=+1$.
Hence, we can simply embed it into the SO(10) adjoint representation of $\bm{45}$. 

In GUTs based on the gauge group SO(10), we can consider a variety of paths
  for the SO(10) breaking down to $G_{\rm SM}$ with various choices of the Higgs representations 
  (see, for example, Ref.~\cite{Ferrari:2018rey}). 
Here we simply consider a set of Higgs fields, ${\bf 45}_H + {\bf 126}_H$.
Under the decomposition of SO(10) $\supset$ SU(5)$\times$U(1)$_X$, 
  ${\bf 45} \supset ({\bf 1},0) \oplus ({\bf 24},0)$ and ${\bf 126} \supset ({\bf 1}, -10)$ \cite{Slansky:1981yr}. 
Thus, the SO(10) gauge group is broken down to 
  ${\rm SO(10)} \to G_{\rm SM} \times {\rm U(1)}_X$ by $\langle {\bf 45}_H \rangle$ 
  and then $G_{\rm SM} \times {\rm U(1)}_X \to G_{\rm SM}$ by $\langle {\bf 126}_H \rangle$.
Since the U(1)$_X$ symmetry broken by $\langle {\bf 126}_H \rangle \supset \langle ({\bf 1}, -10) \rangle$,
  the $Z_2$ parity remains as a remnant symmetry.

As usual in the GUTs, we need to consider mass splittings among the components in a representation ${\bf R}$
  to leave a desired representation under $G_{\rm SM}$ light while the others are heavy. 
For this purpose, we adopt the same idea as the doublet-triplet Higgs mass splitting in the SU(5) GUT model. 
For example, to have a light color octet scalar embedded into ${\bf 45}$, we consider the scalar potential, 
\begin{eqnarray}
  V \supset  m^2_{45} {\rm Tr}\left[{\bf 45} \, {\bf 45}\right]  + \lambda_{45} {\rm Tr}\left[  {\bf 45}_H {\bf 45}_H {\bf 45} \, {\bf 45} \right]. 
\end{eqnarray}
Associated with ${\rm SO(10)} \to G_{\rm SM} \times {\rm U(1)}_X$ by $\langle {\bf 45}_H \rangle$, 
   the second term in the right-hand side generates the mass terms for the SM multiplets in ${\bf 45}$. 
Since SO(10) is broken, the SM multiplets of different representations embedded in ${\bf 45}$ 
   acquire different mass values 
   from the second term through different values of the Clebsch-Gordan coefficients. 
We then fine-tune the common mass of the first term so as to leave only the color-octet scalar light.

Since the MDM is embedded into the $\bm{2640}$ representation which is huge, 
   the beta function coefficient of the GUT gauge coupling $\alpha_X$ is very large. 
 As a result, this gauge coupling blows up in its running right above the GUT scale. 
This is a common problem of SO(10) GUT modes with large representation fields, 
   such as the so-called minimal supersymmetric SO(10) GUT 
   with a pair of ${\bf 126}+\overline{\bf 126}$ representation chiral multiplets \cite{Lee:1994je}. 
In the practical point of view, the problem lies in the discrepancy between the GUT scale 
   and the (reduced) Plank scale, $M_X \ll M_P$, and we have the lack of technology 
   to analyze the evolution of the GUT gauge coupling beyond the perturbation. 
To avoid this problem, we may consider our GUT model in the warped extra-dimension \cite{Fukuyama:2007ph},
   where the cutoff scale of the effective 4-dimensional theory can get down to the GUT scale 
   from the Planck scale due to the warped geometry \cite{Randall:1999ee}.

\section{Conclusions}
\label{sec:conclusion}
The major missing pieces of the SM are the DM candidate and the neutrino masses. 
The MDM scenario is a very simple framework which supplements the SM 
   with a dark matter candidate as a fermion or scalar SU(2)$_L$ multiplet.  
This scenario is very predictable: 
   Once the spin and SU(2)$_L$ representation of an MDM are fixed,     
   the dark matter physics of this scenario is determined by only one free parameter, 
   the mass of MDM particle. 
Since the stability of the MDM particle is not ensured by the SM gauge symmetries, 
   one may think that this scenario is an effective low energy theory 
   of a more fundamental theory which takes place at high energies. 
In this paper, we have considered an ultraviolet completion of the MDM scenario
   to an SO(10) grand unified theory, which is not only an interesting paradigm 
   of unifying all the SM gauge interactions but also a well-motivated framework 
   for naturally generating tiny neutrino masses through the seesaw mechanism. 
We have examined several SU(2)$_L$ multiplets for the fermion/scalar MDM candidate 
   and considered various phenomenological constraints, 
   such as the successful SM gauge coupling unification, 
   the proton stability, and  the direct/indirect DM detection constraints 
   as well as the absolute electroweak vacuum stability. 
We then have singled out the minimal particle content of the MDM scenario at low energies, 
   which satisfies all the constraints. 
In addition to the SM particle content, our MDM scenario includes 
   an SU(2)$_L$ quintet scalar DM with a 9.4 TeV mass 
   and three degenerate color-octet scalars with the mass of 2 TeV. 
Next, we have examined a way to embed our minimal particle content 
   into SO(10) representations, in which a remnant $Z_2$ symmetry after SO(10) breaking
   ensures the stability of the quintet scalar DM. 
We have found the lowest dimensional representation to be ${\bf 2640}$ for the MDM 
   and ${\bf 45}$ for each color-octet scalar.  
Although the color-octet scalars can be produced at the LHC, 
   we have found that their production cross section is a few orders of magnitude
   below the current experimental limit since their couplings with a pair of gluons 
   are induced at the 1-loop level and very much suppressed.



\section*{Acknowledgments}
The work is supported in part by
Grants-in-Aid for Scientific Research from the Japan Society for the Promotion of Science No.~16K05314 (G.C.C.) 
and the United States Department of Energy Grant No.~DE-SC0012447 (N.O.).
K.H. thanks JASSO for financial support to visit University of Alabama where a part of this work was completed.

\renewcommand{\appendixname}{Appendix }
\appendix*

\section{Dimension-5 operator induced by 1-loop corrections}

\begin{figure}
	\centering
	\includegraphics[width=14cm]{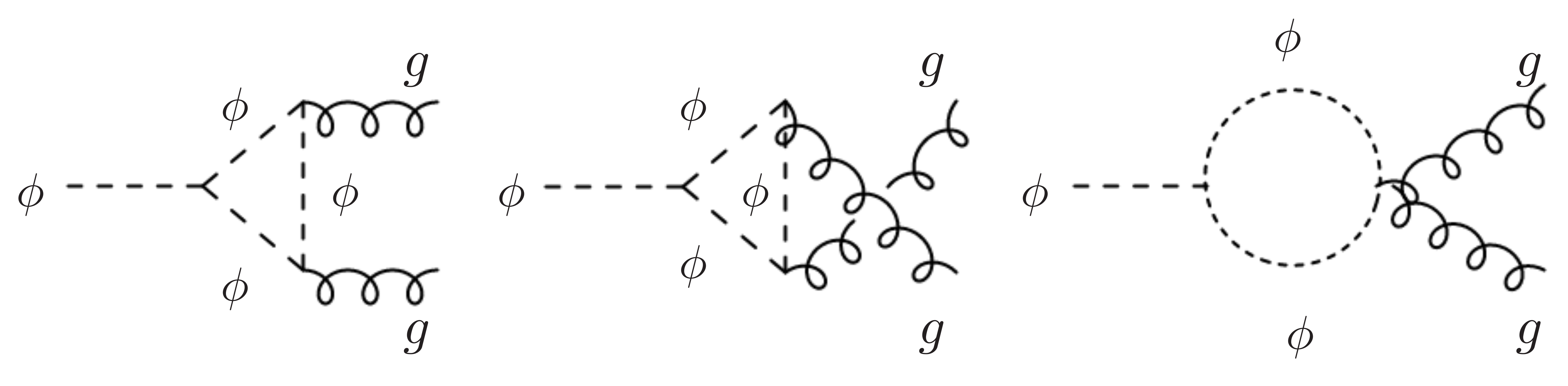}
	\caption{1-loop diagrams inducing the dimension-5 operator}
	\label{DecayDiagram}
\end{figure}%

We parametrize the effective dimension-5 operator as 
	\begin{equation}
		\mathcal{L}_{eff} = \lambda_{eff}^{a b c} \, \phi^a \, G^b_{\mu \nu}  \,  G^{c, \mu \nu},
	\end{equation}
where $\lambda_{eff}^{a b c}$ is the effective coupling constant with a negative mass dimension. 
This effective coupling is induced by quantum corrections at the one-loop level as shown in Fig.~\ref{DecayDiagram}.

The left diagram in Fig.~\ref{DecayDiagram} is calculated to be:
\begin{align}
J_{a b c}^{\mu \nu}=& \int \frac{d^{4} q}{(2 \pi)^{4}} \cdot \frac{i}{q^{2}-m_{\phi}^{2}} \cdot\left(-\frac{g_s}{2}\right) f_{l b n}\left(q+q-k_{2}\right)^{\nu} \cdot \frac{i}{\left(q-k_{2}\right)^{2}-m_{\phi}^{2}} \nonumber \\
& \times\left(-\frac{g_s}{2}\right) f_{n a m}\left[\left(q-k_{2}\right)+\left(q-k_{1}-k_{2}\right)\right]^{\mu} \nonumber \\
& \times \frac{i}{\left(q-k_{1}-k_{2}\right)^{2}-m_{\phi}^{2}} \cdot(i \lambda_3) \mathrm{Tr}\left[T_{m} T_{c} T_{l}\right],
\end{align}
where $q$ is a internal momentum of $\phi$, and $k_{1,2}$ are in-coming momenta for the pair of gluons. 
The coefficient of the loop integral is given by
\begin{equation}
 C = i^4 \lambda_3 \left(-\frac{g_s}{2}\right)^{2} \,  f_{lbn} \, f_{nam} \, \mathrm{Tr} [T_{m}T_c T_l ],
\end{equation}
where $\operatorname{Tr}\left[T^{a} T^{b} T^{c}\right]=\frac{1}{4}(d^{a b c}-i f^{a b c})$.
The integral part is calculated to be
\begin{eqnarray}
I^{\mu \nu} &=& \int \frac{d^{4} q}{(2 \pi)^{4}}  \frac{\left(2 q-k_{1}\right)^{\nu}\left(2 q-k_{1}-2 k_{2}\right)^{\mu}}{\left\{q^{2}-m_{\phi}^{2}\right\}\left\{\left(q-k_{2}\right)^{2}-m_{\phi}^{2}\right\}\left\{\left(q-k_{1}-k_{2}\right)^{2}-m_{\phi}^{2}\right\}} \nonumber \\
&=&((k_1 \cdot k_2)g^{\mu \nu} - k_1^{\nu}k_2^{\mu})I^\prime, 
\end{eqnarray}
with
\begin{eqnarray}
	I^{\prime}&=&8i \int d x dz\int \frac{d^{d} l_{E}}{(2 \pi)^{d}} \frac{(1-x-z) z }{\left(l_{E}^{2}+\Delta\right)^{3}} \nonumber \\
	&=&4 i \int d x d z \frac{1}{(4 \pi)^{2}} \cdot \frac{1}{m_{\phi}^{2}} \frac{x z}{z(z-1)+(1-x-z) z+1} \nonumber \\
	&=&\frac{i}{4 \pi^{2} m_{\phi}^{2}} \int_{0}^{1} d z\left(z-1-\frac{1}{z} \log (1-(1-z) z)\right) \nonumber\\
	&=&\frac{i}{8 \pi^{2} m_{\phi}^{2}}(-1+4 f(1)), 
\end{eqnarray}
where
\begin{equation}
f(t)=\left\{\begin{array}{ll}
{ \left(\arcsin \sqrt{\frac{1}{t}}\right)^{2}} & {(t\geq1)} ,\\
{-\frac{1}{4}\left(\log \frac{1+\sqrt{1-t}}{1-\sqrt{1-t}}\right)} &{(t<1)}.
\end{array} \right.
\end{equation}
In the same way, we calculate the rest of two diagrams in Fig.~\ref{DecayDiagram}.
Summing them all, we obtain the effective coupling: 
\begin{eqnarray}
\lambda_{eff}^{a b c}&=&\lambda_3 \, g_s^{2} \, f_{lbn} \, f_{nam} \, d_{ m c l} \cdot \frac{1}{2^{8} \pi^{2} m_{\phi}^{2}}(-1+4 f(1)) \nonumber \\
&=&\lambda_3 \, \alpha_s \, f_{lbn} \, f_{nam} \, d_{ m c l} \cdot \frac{1}{2^{6} \pi m_{\phi}^{2}}(-1+ \pi^2).
\end{eqnarray}

%
%
%
%

\begin{thebibliography}{99}
\bibitem{Cirelli:2005uq}
M.~Cirelli, N.~Fornengo and A.~Strumia,
``Minimal dark matter,''
Nucl. Phys. B \textbf{753}, 178-194 (2006)
[arXiv:hep-ph/0512090 [hep-ph]].




\bibitem{Cirelli:2007xd}
M.~Cirelli, A.~Strumia and M.~Tamburini,
``Cosmology and Astrophysics of Minimal Dark Matter,''
Nucl. Phys. B \textbf{787}, 152-175 (2007)
[arXiv:0706.4071 [hep-ph]]. 


\bibitem{Cirelli:2009uv}
  M.~Cirelli and A.~Strumia,
 ``Minimal Dark Matter: Model and results,''
  New J.\ Phys.\  {\bf 11}, 105005 (2009)
  [arXiv:0903.3381 [hep-ph]].
  


\bibitem{Minkowski:1977sc}
P.~Minkowski,
Phys. Lett. B \textbf{67}, 421-428 (1977); 
%

\bibitem{Yanagida:1979as}
T.~Yanagida,
``Horizontal gauge symmetry and masses of neutrinos,''
Conf. Proc. C \textbf{7902131}, 95-99 (1979) KEK-79-18-95; 

\bibitem{Gell-Mann:1979vob}
M.~Gell-Mann, P.~Ramond and R.~Slansky,
``Complex Spinors and Unified Theories,''
Conf. Proc. C \textbf{790927}, 315-321 (1979)
[arXiv:1306.4669 [hep-th]]; 


\bibitem{Mohapatra:1979ia}
R.~N.~Mohapatra and G.~Senjanovic,
 ``Neutrino Mass and Spontaneous Parity Nonconservation,''
Phys. Rev. Lett. \textbf{44}, 912 (1980).






\bibitem{Frigerio}M.~Frigerio and T.~Hambye,
  ``Dark matter stability and unification without supersymmetry,''
  Phys.\ Rev.\ D {\bf 81} (2010) 075002
  [arXiv:0912.1545 [hep-ph]].
 
 
 
\bibitem{Aprile:2018dbl}
  E.~Aprile {\it et al.} [XENON Collaboration],
  ``Dark Matter Search Results from a One Ton-Year Exposure of Xenon1T,''
  Phys.\ Rev.\ Lett.\  {\bf 121} (2018) no.11,  111302
  [arXiv:1805.12562 [astro-ph.CO]].
 
 
\bibitem{Dunsky:2020yhv}
D.~Dunsky, L.~J.~Hall and K.~Harigaya,
``Dark Matter Detection, Standard Model Parameters, and Intermediate Scale Supersymmetry,''
JHEP \textbf{04}, 052 (2021)
[arXiv:2011.12302 [hep-ph]].


\bibitem{Okada:2021uqk}
N.~Okada and O.~Seto,
``Superheavy WIMP dark matter from incomplete thermalization,''
Phys. Lett. B \textbf{820}, 136528 (2021)
[arXiv:2103.07832 [hep-ph]].
 
 
 
\bibitem{Griest:1989wd}
K.~Griest and M.~Kamionkowski,
``Unitarity Limits on the Mass and Radius of Dark Matter Particles,''
Phys. Rev. Lett. \textbf{64}, 615 (1990).
 
 

\bibitem{Aghanim:2018eyx}
  N.~Aghanim {\it et al.} [Planck Collaboration],
  ``Planck 2018 Results. Vi. Cosmological Parameters,''
  arXiv:1807.06209 [astro-ph.CO].
  

\bibitem{Bottaro:2021snn}
S.~Bottaro, D.~Buttazzo, M.~Costa, R.~Franceschini, P.~Panci, D.~Redigolo and L.~Vittorio,
``Closing the window on WIMP Dark Matter,''
[arXiv:2107.09688 [hep-ph]].  
  


\bibitem{RGE1}
M.~E.~Machacek and M.~T.~Vaughn,
``Two Loop Renormalization Group Equations In A General Quantum Field Theory.
1. Wave Function Renormalization,''
Nucl.\ Phys.\ B {\bf 222}, 83 (1983); 
%
``Two Loop Renormalization Group Equations In A General Quantum Field Theory.
2. Yukawa Couplings,''
Nucl.\ Phys.\ B {\bf 236}, 221 (1984);
%
``Two Loop Renormalization Group Equations In A General Quantum Field Theory.
3. Scalar Quartic Couplings,''
Nucl.\ Phys.\ B {\bf 249}, 70 (1985). 


\bibitem{RGE2}
C.~Ford, I.~Jack and D.~R.~T.~Jones,
  ``The Standard Model Effective Potential at Two Loops,''
  Nucl.\ Phys.\  {\bf B387} (1992) 373,
  [Erratum-ibid.\  {\bf B504} (1997)  551].
  
\bibitem{RGE3}  
H.~Arason, D.~J.~Castano, B.~Keszthelyi, S.~Mikaelian,
 E.~J.~Piard, P.~Ramond and B.~D.~Wright,
     Phys.\ Rev.\  D {\bf 46}, 3945 (1992). 
     
\bibitem{RGE4}     
V.~D.~Barger, M.~S.~Berger and P.~Ohmann,
  ``Supersymmetric grand unified theories: Two loop evolution of gauge and
  Yukawa couplings,''
  Phys.\ Rev.\  D {\bf 47}, 1093 (1993). 
  
\bibitem{RGE5}      
M.~X.~Luo and Y.~Xiao,
  ``Two-loop renormalization group equations in the standard model,''
  Phys.\ Rev.\ Lett.\  {\bf 90}, 011601 (2003).



\bibitem{RGE_Higgs_quartic}
See, for instance, 
D.~Buttazzo, G.~Degrassi, P.~P.~Giardino, G.~F.~Giudice, F.~Sala, A.~Salvio and A.~Strumia,
 ``Investigating the near-criticality of the Higgs boson,''
JHEP {\bf 1312}, 089 (2013) 
[arXiv:1307.3536 [hep-ph]] and references therein.


 \bibitem{Nishino:2009aa}
  H.~Nishino {\it et al.} 
  ``Search for Proton Decay via $p \to e^+ \pi^0$ and $ p \to \mu^+ \pi^0$ in a Large Water Cherenkov Detector,''
  Phys.\ Rev.\ Lett.\  {\bf 102}, 141801 (2009)
  [arXiv:0903.0676 [hep-ex]].


\bibitem{Hisano:2015rsa}
  J.~Hisano, K.~Ishiwata and N.~Nagata,
  ``QCD Effects on Direct Detection of Wino Dark Matter,''
  JHEP {\bf 1506}, 097 (2015)
  [arXiv:1504.00915 [hep-ph]].


\bibitem{Cuoco:2017iax}
  A.~Cuoco, J.~Heisig, M.~Korsmeier and M.~Kramer,
  ``Constraining heavy dark matter with cosmic-ray antiprotons,''
  JCAP {\bf 1804}, 004 (2018)
  [arXiv:1711.05274 [hep-ph]].
  
  

\bibitem{Aguilar:2016kjl}
  M.~Aguilar {\it et al.} [AMS Collaboration],
  ``Antiproton Flux, Antiproton-to-Proton Flux Ratio, and Properties of Elementary Particle Fluxes in Primary Cosmic Rays Measured with the Alpha Magnetic Spectrometer on the International Space Station,''
  Phys.\ Rev.\ Lett.\  {\bf 117}, no. 9, 091103 (2016).
 

\bibitem{Han:2010rf}
  T.~Han, I.~Lewis and Z.~Liu,
  ``Colored Resonant Signals at the LHC: Largest Rate and Simplest Topology,''
  JHEP {\bf 1012}, 085 (2010)
  [arXiv:1010.4309 [hep-ph]].



\bibitem{CMS:2019gwf}
A.~M.~Sirunyan \textit{et al.} [CMS],
``Search for high mass dijet resonances with a new background prediction method in proton-proton collisions at $\sqrt{s} =$ 13 TeV,''
JHEP \textbf{05}, 033 (2020)
[arXiv:1911.03947 [hep-ex]].


\bibitem{Yamatsu:2015npn}
  N.~Yamatsu,
  ``Finite-Dimensional Lie Algebras and Their Representations for Unified Model Building,''
  arXiv:1511.08771 [hep-ph].

\bibitem{Ferrari:2018rey}
  S.~Ferrari, T.~Hambye, J.~Heeck and M.~H.~G.~Tytgat,
  Phys.\ Rev.\ D {\bf 99}, no. 5, 055032 (2019)
  [arXiv:1811.07910 [hep-ph]].
 

\bibitem{Slansky:1981yr}
R.~Slansky,
``Group Theory for Unified Model Building,''
Phys. Rept. \textbf{79}, 1-128 (1981).


\bibitem{Lee:1994je}
D.~G.~Lee and R.~N.~Mohapatra,
``Automatically R conserving supersymmetric SO(10) models and mixed light Higgs doublets,''
Phys. Rev. D \textbf{51}, 1353-1361 (1995)
[arXiv:hep-ph/9406328 [hep-ph]].


\bibitem{Fukuyama:2007ph}
T.~Fukuyama, T.~Kikuchi and N.~Okada,
``Solving problems of 4D minimal SO(10) model in a warped extra dimension,''
Phys. Rev. D \textbf{75}, 075020 (2007)
[arXiv:hep-ph/0702048 [hep-ph]].



\bibitem{Randall:1999ee}
L.~Randall and R.~Sundrum,
``A Large mass hierarchy from a small extra dimension,''
Phys. Rev. Lett. \textbf{83}, 3370-3373 (1999)
[arXiv:hep-ph/9905221 [hep-ph]].







\end{thebibliography}


\end{document}